\documentclass[12pt]{article}
\usepackage{graphicx}
\usepackage{subfigure,epsfig,epstopdf}
\usepackage{mathptmx,mathrsfs}
\usepackage{amsfonts,amsmath,amssymb,amsthm}
\usepackage{indentfirst}
\usepackage{cite,hyperref}
\usepackage{enumitem}
\usepackage{color}

\numberwithin{equation}{section}
\newcommand{\nn}{\nonumber \\}
\newcommand{\sumint}{\sum \!\!\!\!\!\!\!\! \int }

\DeclareMathAlphabet{\mathcal}{OMS}{cmsy}{b}{n}
\newcommand{\email}[1]{\footnote{E-mail: \href{mailto:#1}{#1}}}

\makeatother

\begin{document}

\title{Thermal effects of a photon gas with a deformed Heisenberg algebra }

\author{R.~Bufalo \email{rodrigo.bufalo@dfi.ufla.br} \\
\textit{ Departamento de F\'{i}sica, Universidade Federal de Lavras,}\\
\textit{ Caixa Postal 3037, 37200-000 Lavras, MG, Brazil} \\
}

\maketitle
\begin{abstract}
In this paper we have consider the thermodynamics of a photon gas subject to the presence of a minimal measurable length following from a covariant extension of the original generalized uncertainty principle (GUP).
After establishing consistently a generalized dynamics, we define a GUP deformed Maxwell invariant  which serves as the basis for our study.
In order to highlight the GUP effects we compute the one- and two-loop order contribution to the partition function at the high-temperature limit.
Afterwards, by computing the internal energy density we conclude that the additional terms can be seen as corrections $\delta\sigma_{\rm{gup}}$ to the Stefan-Boltzmann law due to GUP effects.
\end{abstract}

\newpage
\tableofcontents


\section{Introduction}

\label{sec1}

Along the last decades several heuristic proposals have provided model-independent features and insights for a better understanding of the Nature behaviour at shortest distances, i.e. of a quantum theory of gravity, these are highly motivated with phenomenological inspirations \cite{ref23}.
The search for a common description of particle physics and gravity and for a quantum theory
of the gravitational sector is certainly one of the most outstanding and longstanding problems in physics. 
Space-time noncommutativity and non-Heisenberg uncertainty relations naturally emerges at Plank scale in attempts to accommodate
Quantum Mechanics and General Relativity in a common framework \cite{ref1,ref2,Maggiore:1993kv,doplicher}; it is generally believed that our smooth classical picture of spacetime
should break down at small distances since quantum fluctuations start to dominate.
  
One common feature of such frameworks is the existence of a minimal measure length that is ascribed to 
quantum gravitational effects, or even that the continuum representation of spacetime breaks near to
Planck scale $E_{Pl}$, suggesting in this case that Planck's length $\ell_{Pl}$ acts as a minimal measurable length scale \cite{ref27}
 \footnote{Notice however that the not necessarily the minimal length must be at Planck scale. There are some proposals where the minimal length is in an intermediary scale, placed between the electroweak and Planck scale. }
in almost all frameworks of quantum gravity (string theory, black hole physics, loop quantum gravity, etc)  \cite{ref1,ref2}.

In this way, in order to incorporate the presence of a minimal measurable length scale in a given theory, its canonical structure is changed and hence Heisenberg uncertainty principle is modified, which is then generalized
to a new uncertainty principle, the so-called generalized uncertainty principle (GUP) that encompass this minimal length scale \cite{ref2,ref3,KalyanaRama:2001xd,Hossenfelder:2003jz,Tawfik:2014zca,Tawfik:2015rva}
\begin{equation}
\Delta \hat{x} \gtrsim \frac{\hbar}{\Delta \hat{p}} +\text{const.}~G~\Delta \hat{p}
\end{equation}
Another consequence in order to encompass the presence of a minimal length is that the canonical Heisenberg algebra, $\left[\hat{x}, \hat{p}\right]=i \hbar$, is modified to a non canonical form -- in agreement to the generalized uncertainty principle. Thus, in this context, the commutation relation of the position and momentum operators is now momentum and/or position operator dependent and can be represented generally by the expression  $\left[\hat{x}, \hat{p}\right]=i \hbar f(\hat{x},\hat{p})$, where $f(\hat{x},\hat{p})$ is some function of the operators $\hat{x}$ and $\hat{p}$; so that all the GUP can be obtained from a given function $f(\hat{x},\hat{p})$. Notice that the ordinary case is recovered when this function goes to unit.

In the sense of a GUP, a simple deformation of the Heisenberg algebra given as  \cite{ref1,ref2,Maggiore:1993kv}
\begin{equation}
\left[\hat{x}^{i},\hat{p}_{j}\right]=i\hbar\left(\delta_{j}^{i}+\tilde{\alpha}\left(\hat{p}^{2}\delta_{j}^{i}+2\hat{p}^{i}\hat{p}_{j}\right)\right),\label{eq:0.1}
\end{equation}
is found to be consistent with the existence of a minimum length, with $\tilde{\alpha}=\tilde{\alpha}_{0}\left(\ell_{P}/\hbar\right)^{2}=\tilde{\alpha}_{0}/\left(M_{P}c\right)^{2}$,
and $\alpha_{0}$ is a constant assumed to be of order of unit. 

On the other hand, in a generalization of special relativity, there are approaches that suggest the existence
of an independent observer scale which could be a Planck
energy scale, this is the so-called doubly special relativity (DSR)  \cite{ref12,ref14,ref15}.  
It should be remarked that the interesting thing about DSR is that
it preserves Lorentz symmetry and the basic postulates
of special relativity, but in addition it introduces an upper limit of energy.
It is also possible to express the DSR feature of a maximum momentum scale in the form of a deformed algebra  \cite{ref14,ref15}
\begin{equation}
\left[\hat{x}^{i},\hat{p}_{j}\right]=i\hbar\left(\delta_{j}^{i}-\tilde{\varepsilon}\left(\sqrt{\hat{p}^{2}}\delta_{j}^{i}+\frac{\hat{p}^{i}\hat{p}_{j}}{\sqrt{\hat{p}^{2}}}\right)\right),\label{eq:0.2}
\end{equation}
where $\tilde{\varepsilon}=\ell_{P}$. One can, however, define a new algebra encompassing both features of GUP and DSR, minimal length and maximum momentum, respectively, so that the commutators read \cite{ref20}  
\begin{equation}
\left[\hat{x}^{i},\hat{p}_{j}\right]=i\hbar\left(\delta_{j}^{i}-\varepsilon\left(\sqrt{\hat{p}^{2}}\delta_{j}^{i}+\frac{\hat{p}^{i}\hat{p}_{j}}{\sqrt{\hat{p}^{2}}}\right)+\alpha\left(\hat{p}^{2}\delta_{j}^{i}+2\hat{p}^{i}\hat{p}_{j}\right)\right),\label{eq:0.3}
\end{equation}
in this case the relations $\left[\hat{x}^{i},\hat{x}_{j}\right]=0=\left[\hat{p}^{i},\hat{p}_{j}\right]$
are ensured  via Jacobi identity. For instance, the one-dimensional GUP is found to be
\begin{align}
\Delta\hat{x}\Delta\hat{p}   \geq\frac{\hbar}{2}\left(1-2\varepsilon\left\langle \hat{p}\right\rangle +\left(3\alpha-2\varepsilon^{2}\right)\left\langle \hat{p}^{2}\right\rangle \right), 
\end{align}
in particular, we see that the choice $\alpha=2\varepsilon^{2}$ reproduces
the results of the GUP as proposed by ref.~\cite{ref20}. As a result, we find out from this expression that
$\Delta\hat{x}\geq\left(\Delta\hat{x}\right)_{\rm{min}}\approx\varepsilon_{0}\ell_{P}$
and $\Delta\hat{p}\leq\left(\Delta\hat{p}\right)_{\rm{max}}\approx\frac{M_{P}c}{\varepsilon_{0}}$.
Additionally, we see that the physical momentum is 	modified as follows
\begin{equation}
x_i = \tilde{x}_i , \quad p_i = \tilde{p}_i 	\left( 1-\alpha \sqrt{\tilde{p}^2}+2\alpha^2 \tilde{p}^2	\right),
\end{equation} 
where the tilde operators satisfy the usual canonical Heisenberg algebra. In view of these effects,
one observes a modification into the energy-momentum dispersion relation so that $E^2(p)=p^2c^2 \left(1-\alpha \sqrt{p^2}\right)^2+m^2c^4$.
Such modifications definitely have prominent effects in a plethora of quantum physical phenomena \cite{ref21,Das:2009hs,Majhi:2013koa,ref22,Pramanik:2014mma}. 
One particular and rich context  where GUP implies essential effects is into statistical and
thermodynamic properties of any physical system  \cite{Arcioni:1999hw, Nozari:2006gg,Das:2010gk,Chandra:2011nj,Ali:2014dfa,ref36}. 
This is due to the fact that GUP changes  the number of accessible
microscopic states of the phase space volume, which thus modify the density states.

Although studies have been presented investigating thermal effects on physical systems, ideal gas or photon gas, in the presence of different types of GUP, we wish here to analyse photon gas thermodynamics from a field theoretical point-of-view.
Hence, for this purpose we will consider a covariant extension of the algebra \eqref{eq:0.3} proposed in Ref.~\cite{Faizal:2014pia}, and employ it in the analysis of thermal effects on a photon gas.
The work is organized as follows. In Sec.~\ref{sec2}, we briefly review the
main aspects of the covariant deformed Heisenberg algebra, and deduce a generalized field strength tensor associated
to this deformed algebra.
Moreover, we propose a functional action for the gauge field based on this generalized field strength tensor,
and determine the respective Feynman rules which takes into account leading GUP effects.
In Sec.~\ref{sec3}, we compute the one-loop diagrams contribution to the effective Lagrangian.
Next, in Sec.~\ref{sec4}, we proceed and calculate the two-loop graphs contribution at high-temperature limit to the effective Lagrangian. Thus, based on the one- and two-loop results we compute the internal energy density.  
Finally, our conclusions and remarks are given in Sec.~\ref{sec5}.
  

\section{Covariant deformed Heisenberg algebra}

\label{sec2}

It is of our interest to consider a GUP formulated so that the time component
is included \cite{Faizal:2014pia} -- an extension of the algebra \eqref{eq:0.3}. Thus, we have
\begin{equation}
\left[\hat{x}^{\mu},\hat{p}_{\nu}\right]=i\left(\delta_{\nu}^{\mu}-\varepsilon\left(\sqrt{\hat{p}^{2}}\delta_{\nu}^{\mu}+\frac{\hat{p}^{\mu}\hat{p}_{\nu}}{\sqrt{\hat{p}^{2}}}\right)+\alpha\left(\hat{p}^{2}\delta_{\nu}^{\mu}+2\hat{p}^{\mu}\hat{p}_{\nu}\right)\right).\label{eq:1.1}
\end{equation}
Moreover, we can define a new set of phase-space variables
\begin{equation}
\hat{x}^{\mu}=x^{\mu},\quad\hat{p}_{\nu}=\left(1-\varepsilon\sqrt{p^{2}}+\alpha p^{2}\right)p_{\nu},\label{eq:1.2}
\end{equation}
so that they satisfy the canonical commutation relations
$\left[x^{\mu},p_{\nu}\right]=i$, it can be shown that \eqref{eq:1.1} is satisfied.
In particular, $p_{\nu}$ can be interpreted as the momentum at low
energies (where the representation in position space reads $p_{\nu}=-i\partial_{\nu}$)
while $\hat{p}_{\nu}$ is that at higher energies.

Although this generalized GUP has both minimal length and maximal momentum (time
and energy) the $\varepsilon$-dependent term gives non-local contributions
which complicate substantially our analysis in regard to the gauge
fields (for instance, when taking the minimal coupling in configuration
representation $p_{\nu}=-i\partial_{\nu}\rightarrow-i\nabla_{\nu}$,
we obtain $\sqrt{p^{2}}\rightarrow\sqrt{-\nabla_{\nu}\nabla^{\nu}}=\sqrt{-\left(\partial+iA\right)_{\nu}\left(\partial+iA\right)^{\nu}}$).
Hence, we shall take $\varepsilon=0$ and concentrate our analysis only
on the $\alpha$-dependent terms -- that will engender minimal length and time effects.

Perhaps an alternative way to workaround the problematic non-local behavior due to the use of
minimal coupling in the above realization of position and momentum operators, i.e. those $\sqrt{p^{2}}$ $\varepsilon$-dependent terms, is to find a generalized procedure of minimal coupling such as in Ref.~\cite{Harikumar:2011um}. There the authors find Lorentz force and Maxwell's equations on kappa-Minkowski space-time by postulating that the momentum (with gauge field) $p\rightarrow\pi$ satisfies the same commutation relation as $p$.
Hence, this generalized approach could help us to circumvent the above non-local issues, or at least give us insights, since it is expected that this approach give us different realization for the operators $\hat{x}^{\mu}$ and $\hat{p}^{\mu}$ in \eqref{eq:1.2}, so that terms like $\sqrt{p^{2}}$ are fully removed, but at the same preserving the GUP structure. Nonetheless, this proposal should be further elaborated and in a positive case analyzed for a development of a deformed electrodynamics.

Based on the above discussion, one can immediately conclude that a GUP, actually the variables from \eqref{eq:1.2}, leads to an action of infinite
order in derivatives, implying thus in an infinite series of interaction terms.
This is the most interesting point that we wish to explore.
In particular, one can easily see that a calculation to the first
order already yields additional interactions terms of the gauge field.

We shall now apply GUP, i.e. \eqref{eq:1.2}, to the free Dirac action and then resort to
the (minimal coupling) gauge principle in order to develop the formalism for the gauge
fields. In this deformed scenario, the action reads   \cite{Faizal:2014pia2}
\begin{equation}
S=\int d^{4}x\overline{\psi}\left[i\left(1-\alpha\partial^{\lambda}\partial_{\lambda}\right)\gamma^{\mu}\partial_{\mu}-m\right]\psi,
\end{equation}
it is easy to see that this action is invariant under a certain global symmetry transformation.
Now, if we extend this to a local transformation, the additional derivatives
also act on the local unitary operator $U\left(x\right)$, so we must have the replacement \cite{ref30}
\begin{equation}
\left(1-\alpha\partial^{\lambda}\partial_{\lambda}\right)\partial_{\mu}\rightarrow\left(1-\alpha\nabla^{\lambda}\nabla_{\lambda}\right)\nabla_{\mu}.\label{eq:1.3}
\end{equation}
Hence, from this minimal coupling, in the same way the usual covariant derivative satisfies 
\begin{equation}
\delta\left(\nabla_{\mu}\psi\right)=U\left(x\right)\left(\nabla_{\mu}\psi\right)
\end{equation}
one can show that the transformation of the subsequent term reads
\begin{equation}
\delta\left(\nabla^{\lambda}\nabla_{\lambda}\left(\nabla_{\mu}\psi\right)\right)=U\left(x\right)\left(\nabla^{\lambda}\nabla_{\lambda}\left(\nabla_{\mu}\psi\right)\right),
\end{equation}
note that $\left[\gamma^{\mu},U\left(x\right)\right]=0$, since here
$U\left(x\right)$ does not refers to a spacetime transformation,
otherwise we should have $\gamma^{\mu}\rightarrow U\left(x\right)\gamma^{\mu}U^{\dagger}\left(x\right)$.
According to the above transformation rules, it shows to be convenient
to define a GUP covariant derivative 
\begin{equation}
\mathcal{D}_{\mu}=\left(1-\alpha\nabla^{\lambda}\nabla_{\lambda}\right)\nabla_{\mu},\label{eq:1.4}
\end{equation}
in particular, we see that it behaves as the usual covariant derivative under
a local gauge transformation, i.e. $\mathcal{D}_{\mu}\rightarrow U\left(x\right)\mathcal{D}_{\mu}U^{\dagger}\left(x\right)$.

Consequently, a generalized definition of the field strength tensor
for the gauge fields follows naturally. This is achieved by the definition
\begin{equation}
i\mathcal{F}_{\mu\nu}\Phi=\left[\mathcal{D}_{\mu},\mathcal{D}_{\nu}\right]\Phi.\label{eq:1.5}
\end{equation}
In particular, this relation is well motivated and necessary so
that the field equations following from the action for the gauge field
(built from such definition) contain GUP effects. Furthermore, notice that from this definition the gauge invariance of the field strength tensor follows as $\mathcal{F}_{\mu\nu}\rightarrow U\left(x\right)\mathcal{F}_{\mu\nu} U^{\dagger}\left(x\right)$.

We can now compute an explicit expression for the generalized field
strength tensor up to $\alpha$-order, so that it yields
\begin{align}
i\mathcal{F}_{\mu\nu}\Phi   & =i\bigg(\left(1-2\alpha\nabla_{\rho}\nabla^{\rho}\right)F_{\mu\nu}-\alpha\left(\nabla^{\rho}F_{\mu\rho}+F_{\mu\rho}\nabla^{\rho}\right)\nabla_{\nu} \nn 
& -\alpha\left(F_{\rho\nu}\nabla^{\rho}+\nabla^{\rho}F_{\rho\nu}\right)\nabla_{\mu}\bigg)\Phi+\mathcal{O}\left(\alpha^{2}\right),\label{eq:1.6}
\end{align}
here we have introduced the usual Abelian field strength $F_{\mu\nu}= \partial_\mu A_\nu -\partial_\nu A_\mu$. Thus we have that the Maxwell's invariant reads \cite{ref30,Dias:2016lkg}
\begin{align}
\int\mathcal{F}_{\mu\nu}\mathcal{F}^{\mu\nu} & =\int\Bigl\{ F^{\mu\nu}F_{\mu\nu}+4\alpha\left(\partial^{\rho}F_{\mu\nu}\right)^{2} \nn
&-8i\alpha\left(A_{\rho}F^{\rho\nu}\right)\left(\partial^{\mu}F_{\mu\nu}\right)+8\alpha\left(A^{\mu}F_{\mu\nu}\right)^{2}+4\alpha\left(A^{\rho}F^{\mu\nu}\right)^{2}\Bigr\}+\mathcal{O}\left(\alpha^{2}\right).\label{eq:1.7}
\end{align}
The gauge invariance of the generalized Maxwell's invariant follows from the gauge invariance of the field strength tensor $\mathcal{F}_{\mu\nu}$.
Moreover, notice that the first two terms are those present in the Bopp-Podolsky
generalized electrodynamics \cite{ref35,ref39}. These are higher-derivative (HD) terms and they have several
problems associated with their presence, for instance unitarity \cite{ref37}. However, they have prominent role in gravity \cite{ref38}.
Nonetheless, we shall use the action defined by \eqref{eq:1.7} in our analysis of thermal effects.

On the other hand, we can also consider a second invariant
\begin{align}
\int\mathcal{F}_{\mu\nu}\mathcal{G}^{\mu\nu} & =\int\Bigl\{ G^{\mu\nu}F_{\mu\nu}+4\alpha\partial_{\theta}G^{\mu\nu}\partial^{\theta}F_{\mu\nu}+4\alpha G^{\mu\nu}F_{\mu\nu}A_{\theta}A^{\theta}+8\alpha A^{\theta}F_{\mu\theta}A_{\nu}G^{\mu\nu}\nonumber \\
 & +4i\alpha A^{\theta}F_{\mu\nu}\partial_{\theta}G^{\mu\nu}-4i\alpha G^{\mu\nu}A_{\theta}\partial^{\theta}F_{\mu\nu}+8i\alpha A^{\theta}F_{\theta\mu}\partial_{\nu}G^{\mu\nu}\Bigr\}+O\left(\alpha^{2}\right)
\end{align}
which is also gauge invariant, where we have introduced a dual for the field strength tensor $\mathcal{F}_{\mu\nu}$ defined as field strength tensor $\mathcal{G}_{\mu\nu}=\frac{1}{2}\epsilon_{\mu \nu \lambda\sigma}\mathcal{F}^{\lambda\sigma}$.  
That can be used for instance in the analysis of GUP effects on nonlinear electrodynamics, for instance in the Born-Infeld electrodynamics.

In order to proceed with our analysis let us consider the following functional action 
\begin{align}
\mathcal{S}_{c} & =-\frac{1}{4}\int\mathcal{F}_{\mu\nu}\mathcal{F}^{\mu\nu}\label{eq:1.14}
\end{align}
The next step is to derive the Feynman rules from \eqref{eq:1.14}. The gauge field propagator is found when we choose a suitable gauge fixing condition, in which we consider here a non-mixing Lorenz condition given as  \cite{ref24,ref25}
\begin{equation}
\Omega\left[A\right]=\sqrt{\left(1+\alpha\square\right)}\partial_{\mu}A^{\mu}=0,\label{eq:1.15}
\end{equation}
notice that this is a pseudodifferential operator \cite{ref26}. Thus, we have that the propagator reads 
\begin{equation}
iD^{\mu\nu}=\frac{1}{\left(1-\alpha k^{2}\right)k^{2}}\eta^{\mu\nu}-\left(1-\xi\right)\frac{1}{\left(1-\alpha k^{2}\right)k^{4}}k^{\mu}k^{\nu},\label{eq:1.16}
\end{equation}
In particular, we take the Feynman gauge $\xi=1$, so that we find a simple expression 
\begin{equation}
iD^{\mu\nu}=\frac{1}{\left(1-\alpha k^{2}\right)k^{2}}\eta^{\mu\nu}=\left[\frac{1}{k^{2}}-\frac{1}{\left(k^{2}-\alpha^{-1}\right)}\right]\eta^{\mu\nu}.\label{eq:1.17}
\end{equation}

The remaining Feynman rules, i.e. for three and four gauge fields,
can be obtained straightforwardly from the action \eqref{eq:1.14}.
The three-gauge field vertex  $\left\langle A_{\mu}A_{\nu}A_{\lambda}\right\rangle $ reads
\begin{align}
i\Gamma^{\mu\nu\lambda}\left(p,q,k\right) & =-\frac{\alpha}{2}\delta\left(p+q+k\right)\biggl[\left(q^{\mu}\delta_{\sigma}^{\nu}+p^{\nu}\delta_{\sigma}^{\mu}-\eta^{\nu\mu}\left(p_{\sigma}+q_{\sigma}\right)\right)\left(k^{2}\eta^{\lambda\sigma}-k^{\lambda}k^{\sigma}\right)\nn
 & +\left(k^{\mu}\delta_{\sigma}^{\lambda}+p^{\lambda}\delta_{\sigma}^{\mu}-\eta^{\lambda\mu}\left(p_{\sigma}+k_{\sigma}\right)\right)\left(q^{2}\eta^{\nu\sigma}-q^{\nu}q^{\sigma}\right)\nn
 &+\left(k^{\nu}\delta_{\sigma}^{\lambda}+q^{\lambda}\delta_{\sigma}^{\nu}-\eta^{\lambda\nu}\left(q_{\sigma}+k_{\sigma}\right)\right)\left(p^{2}\eta^{\mu\sigma}-p^{\mu}p^{\sigma}\right)\biggr],\label{eq:A.1}
\end{align}
while the four-gauge field vertex $\left\langle A_{\mu}A_{\nu}A_{\lambda}A_{\sigma}\right\rangle $ is
\begin{align}
& i\Gamma^{\mu\nu\lambda\sigma}\left(p,q,k,r\right) = \nn
& =\frac{\alpha}{4}\left(2\eta^{\psi\pi}\eta^{\phi\tau}\eta^{\beta\omega}+\eta^{\phi\pi}\eta^{\psi\tau}\eta^{\beta\omega}\right)\delta\left(p+q+k+r\right)\nn
 & \times\biggl\{\left(r_{\phi}\delta_{\beta}^{\sigma}-r_{\beta}\delta_{\phi}^{\sigma}\right)\left(q_{\pi}\delta_{\omega}^{\nu}-q_{\omega}\delta_{\pi}^{\nu}\right)\left(\delta_{\tau}^{\mu}\delta_{\psi}^{\lambda}+\delta_{\psi}^{\mu}\delta_{\tau}^{\lambda}\right)+\left(r_{\phi}\delta_{\beta}^{\sigma}-r_{\beta}\delta_{\phi}^{\sigma}\right)\left(k_{\pi}\delta_{\omega}^{\lambda}-k_{\omega}\delta_{\pi}^{\lambda}\right)\left(\delta_{\tau}^{\mu}\delta_{\psi}^{\nu}+\delta_{\psi}^{\mu}\delta_{\tau}^{\nu}\right)\nn
 & +\left(r_{\phi}\delta_{\beta}^{\sigma}-r_{\beta}\delta_{\phi}^{\sigma}\right)\left(p_{\pi}\delta_{\omega}^{\mu}-p_{\omega}\delta_{\pi}^{\mu}\right)\left(\delta_{\tau}^{\nu}\delta_{\psi}^{\lambda}+\delta_{\psi}^{\nu}\delta_{\tau}^{\lambda}\right)\nn
 & +\left(k_{\phi}\delta_{\beta}^{\lambda}-k_{\beta}\delta_{\phi}^{\lambda}\right)\left(q_{\pi}\delta_{\omega}^{\nu}-q_{\omega}\delta_{\pi}^{\nu}\right)\left(\delta_{\psi}^{\sigma}\delta_{\tau}^{\mu}+\delta_{\tau}^{\sigma}\delta_{\psi}^{\mu}\right)+\left(k_{\phi}\delta_{\beta}^{\lambda}-k_{\beta}\delta_{\phi}^{\lambda}\right)\left(r_{\pi}\delta_{\omega}^{\sigma}-r_{\omega}\delta_{\pi}^{\sigma}\right)\left(\delta_{\tau}^{\mu}\delta_{\psi}^{\nu}+\delta_{\psi}^{\mu}\delta_{\tau}^{\nu}\right)\nn
 & +\left(k_{\phi}\delta_{\beta}^{\lambda}-k_{\beta}\delta_{\phi}^{\lambda}\right)\left(p_{\pi}\delta_{\omega}^{\mu}-p_{\omega}\delta_{\pi}^{\mu}\right)\left(\delta_{\tau}^{\nu}\delta_{\psi}^{\sigma}+\delta_{\psi}^{\nu}\delta_{\tau}^{\sigma}\right)\nn
 & +\left(q_{\phi}\delta_{\beta}^{\nu}-q_{\beta}\delta_{\phi}^{\nu}\right)\left(k_{\pi}\delta_{\omega}^{\lambda}-k_{\omega}\delta_{\pi}^{\lambda}\right)\left(\delta_{\psi}^{\sigma}\delta_{\tau}^{\mu}+\delta_{\tau}^{\sigma}\delta_{\psi}^{\mu}\right)+\left(q_{\phi}\delta_{\beta}^{\nu}-q_{\beta}\delta_{\phi}^{\nu}\right)\left(r_{\pi}\delta_{\omega}^{\sigma}-r_{\omega}\delta_{\pi}^{\sigma}\right)\left(\delta_{\tau}^{\mu}\delta_{\psi}^{\lambda}+\delta_{\psi}^{\mu}\delta_{\tau}^{\lambda}\right)\nn
 & +\left(q_{\phi}\delta_{\beta}^{\nu}-q_{\beta}\delta_{\phi}^{\nu}\right)\left(p_{\pi}\delta_{\omega}^{\mu}-p_{\omega}\delta_{\pi}^{\mu}\right)\left(\delta_{\psi}^{\sigma}\delta_{\tau}^{\lambda}+\delta_{\psi}^{\lambda}\delta_{\tau}^{\sigma}\right)\nn
 & +\left(p_{\phi}\delta_{\beta}^{\mu}-p_{\beta}\delta_{\phi}^{\mu}\right)\left(k_{\pi}\delta_{\omega}^{\lambda}-k_{\omega}\delta_{\pi}^{\lambda}\right)\left(\delta_{\psi}^{\sigma}\delta_{\tau}^{\nu}+\delta_{\psi}^{\nu}\delta_{\tau}^{\sigma}\right)+\left(p_{\phi}\delta_{\beta}^{\mu}-p_{\beta}\delta_{\phi}^{\mu}\right)\left(r_{\pi}\delta_{\omega}^{\sigma}-r_{\omega}\delta_{\pi}^{\sigma}\right)\left(\delta_{\tau}^{\nu}\delta_{\psi}^{\lambda}+\delta_{\psi}^{\nu}\delta_{\tau}^{\lambda}\right)\nn
 & +\left(p_{\phi}\delta_{\beta}^{\mu}-p_{\beta}\delta_{\phi}^{\mu}\right)\left(q_{\pi}\delta_{\omega}^{\nu}-q_{\omega}\delta_{\pi}^{\nu}\right)\left(\delta_{\psi}^{\sigma}\delta_{\tau}^{\lambda}+\delta_{\psi}^{\lambda}\delta_{\tau}^{\sigma}\right)\biggr\}.\label{eq:A.2}
\end{align}

It is of phenomenological interest to highlight the effects of these new two tree vertices $\alpha$--coupling in comparison to the usual fermions and photon $e$--coupling.
These two contributions can be in principle compared at the light-by-light scattering, since all three these vertex have a finite contribution.
It is known that at one-loop this process is of $e^4$-order, but it also have $\alpha^4$-order and $\alpha^2$-order contributions from the $\left\langle A_{\mu}A_{\nu}A_{\lambda}\right\rangle $
and  $\left\langle A_{\mu}A_{\nu}A_{\lambda}A_{\sigma}\right\rangle $ vertices, respectively.
But, since the value of $\alpha$ is presumably related to Planck scale, its contribution is rather small
in comparison to the $e$--coupling contribution, giving hence a rather tiny contribution to any outcome value comparable to data.

\section{One-loop calculation}
\label{sec3}

\begin{figure*}[tbp]
\begin{center}
\includegraphics[scale=0.25]{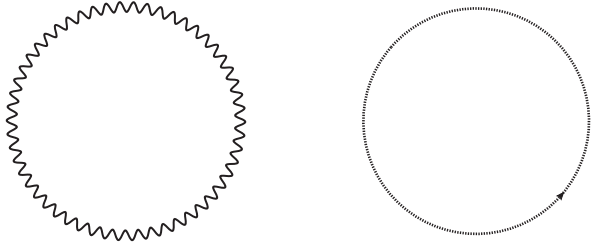}
\end{center}
\caption{ One-loop ring contributions: photon and ghost loops.}
\label{graph1}
\end{figure*}

The first contribution comes from the quadratic part of the action \eqref{eq:1.14}, augmented by
the non-mixing gauge-fixing and ghost fields is given by \footnote{By means of notation, we shall consider henceforth $m^2=\alpha^{-1}$.}
\begin{align}
\mathcal{L}_{AA+cc} & =-\frac{1}{4}F^{\mu\nu}F_{\mu\nu}-\frac{1}{4m^2}\left(\partial^{\rho}F_{\mu\nu}\right)^{2}-\frac{1}{2\xi}\left(\sqrt{\left(1+m^{-2}\square\right)}\partial_{\mu}A^{\mu}\right)^{2}\nn
 & +\partial_{\mu}\overline{c}\sqrt{\left(1+m^{-2}\square\right)}\partial_{\mu}c
\end{align}
The lowest order contributions to the effective Lagrangian are the (one-loop)
ring diagrams from the photon loop and ghost loop Fig.~\ref{graph1}
\begin{equation}
\mathcal{L}^{\left(1\right)}=\mathcal{L}_{\rm{p}} +\mathcal{L}_{\rm{gh}} ,\label{eq 0.4}
\end{equation}
where each contribution reads
\begin{align}
\mathcal{L}_{\rm{ph}} & =-\frac{1}{2}\ln\rm{Det}\left(M_{\mu\nu}\right),\label{eq 0.6}\\
\mathcal{L}_{\rm{gh}} & =\ln\det\left(\sqrt{1+m^{-2}\square}\square\right),\label{eq 0.7}
\end{align}
in which we have the following differential operator 
\begin{equation}
M_{\mu\nu}\left(x,y\right)=\left[\eta_{\mu\nu}\square-\left\{ 1-\frac{1}{\xi}\right\} \partial_{\mu}\partial_{\nu}\right]\left(1+m^{-2}\square\right)\delta\left(x,y\right).\label{eq 0.8}
\end{equation}

Notice that in $\mathcal{L}_{\rm{ph}}$ we also have the determinant
on the spacetime indices in addition to the Hilbert space. Now, in
this case we find the result $\rm{Det}\left(M_{\mu\nu}\right)=\frac{1}{\xi}\left[\left(1+\alpha\square\right)\square\right]^{4}$
\begin{align}
\mathcal{L}_{\rm{ph}} & =-2\ln\det\left(\left(1+m^{-2}\square\right)\square\right),\\
\mathcal{L}_{\rm{gh}} & =\ln\det\left(\sqrt{1+m^{-2}\square}\square\right).
\end{align}
Moreover, in the imaginary time formalism we can express these contributions as 
\begin{align}
\mathcal{L}^{\left(1\right)} & =-\frac{3}{2}\ln\det\left(1+\alpha\square\right)-\ln\det\square,\label{eq 0.11}\\
 & =-\frac{3}{2\beta}\underset{n_{B}}{\sum}\int\frac{d^{\omega-1}p}{\left(2\pi\right)^{\omega-1}}\ln\left[\beta^{2}\left(1-\frac{p^{2}}{m^{2}}\right)\right]-\frac{1}{\beta}\underset{n_{B}}{\sum}\int\frac{d^{\omega-1}p}{\left(2\pi\right)^{\omega-1}}\ln\left[-\beta^{2}p^{2}\right],
\end{align}
where we are considering by means of generality a $\omega$-dimensional spacetime in order to compute the momentum sum/integral. It should emphasized that the sum is over $p_{0}=i\omega_{n_{B}}$, where $\omega_{n_{B}}=\frac{2n\pi}{\beta}$ is the bosonic Matsubara frequency.
In order to evaluate the sum/integrals in Eq.\eqref{eq 0.11}
\begin{align}
\mathcal{L}^{\left(1\right)} & =-\frac{3}{2\beta}\underset{n_{B}}{\sumint  } \ln\left[\beta^{2}\left(1-\frac{\left[\left(q_{0}\right)^{2}-\left\vert q\right\vert ^{2}\right]}{m^{2}}\right)\right] -\frac{1}{\beta}\underset{n_{B}}{\sumint  } \ln\left[-\beta^{2}\left[\left(q_{0}\right)^{2}-\left\vert q\right\vert ^{2}\right]\right].\label{eq 0.19}
\end{align}
where we have introduced the notation for the bosonic sum/integral
\begin{equation}
\underset{n_{B}}{\sumint  } \equiv \underset{n_{B}}{\sum  } \int \frac{d^{\omega -1}p}{\left( 2\pi \right) ^{\omega -1}}.
\end{equation}
It is important to notice that the massive sector has the expected
correct number of three degrees-of-freedom (d.o.f.), this matches the obtained results from Podolsky's
theory \cite{Bonin:2009je}. The massless part has two degrees of freedom ($\# \rm{d.o.f.}/2=2/2=1$),
while the massive sector has three degrees of freedom ($\#\rm{d.o.f.}/2=3/2$).

We should remark that as usual all temperature-independent parts of \eqref{eq 0.19}
lead to a divergent result, i.e., the zero-point energy of the vacuum,
and they are subtracted off since they adds to an unobservable constant.
Next, the massless bosonic sum/integral can be readily evaluated
\begin{align}
I_{B_{1}}^{\left(1\right)}= & \underset{n_{B}}{\sumint  } \ln\left(-\beta^{2}\left[\left(q_{0}\right)^{2}-\left\vert \vec{q}\right\vert ^{2}\right]\right)=2\int\frac{d^{\omega-1}q}{\left(2\pi\right)^{\omega-1}}\ln\left(1-e^{-\beta\omega_{q}}\right),
\end{align}
with $\omega_q = \left\vert \vec{q}\right\vert$. Besides, we can make use of the known result for the bosonic integration
\begin{equation}
\int_{0}^{\infty}\frac{z^{x-1}}{e^{z}-1}dz=\Gamma\left(x\right)\zeta\left(x\right), \label{eq 0.22}
\end{equation}
in order to get 
\begin{equation}
I_{B_{1}}^{\left(1\right)}=-\frac{2\beta^{1-\omega}}{\left(4\pi\right)^{\frac{\omega-1}{2}}}\frac{\Gamma\left(\omega\right)\zeta\left(\omega\right)}{\Gamma\left(\frac{\omega+1}{2}\right)}.\label{eq 0.23}
\end{equation}
Moreover, in order to compute the massive bosonic sum/integral we write
\begin{align*}
I_{B_{2}}^{\left(1\right)}= & \underset{n_{B}}{\sumint  } \ln\left[\beta^{2}\left(1-\frac{\left[\left(q_{0}\right)^{2}-\left\vert q\right\vert ^{2}\right]}{m^{2}}\right)\right]=2\int\frac{d^{\omega-1}q}{\left(2\pi\right)^{\omega-1}}\ln\left(1-e^{-\beta\omega_{M}}\right),
\end{align*}
where $\omega_{m}^{2}=\left\vert q\right\vert ^{2}+m^{2}$. Besides, we can rewrite the above expression as
\begin{align}
I_{B_{2}}^{\left(1\right)}= & -\frac{2\beta}{\left(4\pi\right)^{\frac{\omega-1}{2}}}\frac{1}{\Gamma\left(\frac{\omega+1}{2}\right)}\int_{0}^{\infty}\frac{q^{\omega}}{e^{\beta\sqrt{q^{2}+m^{2}}}-1}\frac{dq}{\sqrt{q^{2}+m^{2}}},
\end{align}
In particular, it is useful to consider the identity $\frac{e^{-\beta\sqrt{q^{2}+m^{2}}}}{1-e^{-\beta\sqrt{q^{2}+m^{2}}}} =\underset{k=1}{\sum}e^{-k\beta\sqrt{q^{2}+m^{2}}}$,
this relation holds since $\beta\sqrt{q^{2}+m^{2}}>0$, thus $e^{-\beta\sqrt{q^{2}+m^{2}}}<1$.
Moreover, by means of a change of variables $z=\sqrt{q^{2}+m^{2}}$ and introducing $z=mw$, we find
\begin{equation}
I_{B_{2}}^{\left(1\right)}=-\frac{2\beta}{\left(4\pi\right)^{\frac{\omega-1}{2}}}\frac{m^{\omega}}{\Gamma\left(\frac{\omega+1}{2}\right)}\underset{k=1}{\sum}\int_{1}^{\infty}dw\left(w^{2}-1\right)^{\frac{\omega-1}{2}}e^{-k\beta mw}.
\end{equation}
We can then make use of the following representation of the modified Bessel function of the second-kind \cite{gradshteyn}
\begin{equation}
K_{n}\left(z\right)=\frac{\sqrt{\pi}}{\Gamma\left(n+\frac{1}{2}\right)}\left(\frac{z}{2}\right)^{n}\int_{1}^{\infty}dx\left(x^{2}-1\right)^{n-\frac{1}{2}}e^{-zx},
\end{equation}
and by recognizing $n=\frac{\omega}{2}$ and $z=k\beta M$, one finds 
\begin{equation}
\frac{1}{\Gamma\left(\frac{\omega+1}{2}\right)}\int_{1}^{\infty}dx\left(x^{2}-1\right)^{\frac{\omega-1}{2}}e^{-k\beta mx}=\frac{1}{\sqrt{\pi}}\left(\frac{2}{k\beta m}\right)^{\frac{\omega}{2}}K_{\frac{\omega}{2}}\left(k\beta m\right).
\end{equation}
Hence, this result allows us to write the final expression for the massive contribution as
\begin{equation}
I_{B_{2}}^{\left(1\right)} =-\frac{4\beta m^{\omega}}{\left(2\pi\right)^{\frac{\omega}{2}}}\underset{k=1}{\sum}\left(\frac{1}{k\beta m}\right)^{\frac{\omega}{2}}K_{\frac{\omega}{2}}\left(k\beta m\right).\label{eq 0.24}
\end{equation}
Therefore, with the results \eqref{eq 0.23} and \eqref{eq 0.24}
we find the complete bosonic contribution \eqref{eq 0.19}
\begin{align}
\mathcal{L}^{\left(1\right)} & =\frac{1}{\beta^{4}}\left[\frac{\pi^{2}}{45}+\frac{3}{16\pi^{2}}\underset{k=1}{\sum}\left(\frac{2\beta m}{k}\right)^{2}K_{2}\left(k\beta m\right)\right].\label{eq 0.25}
\end{align}

Although we have obtained a closed form expression, there is not known
a form for the above series, which means that we can resort to thermal
properties in order to find a suitable approximation for its evaluation \cite{Dolan:1973qd}. 
We can then assume the high-temperature limit, i.e. the inequality holds $\beta m\ll1$, which means
that the parameter $m$ should be much less than the thermal energy.
Thus, we may use the asymptotic expansion for $\left\vert z\right\vert \rightarrow 0 $ \cite{gradshteyn}
\begin{equation}
K_{2}\left(z\right) \sim  \frac{2}{z^2}-\frac{1}{2} + \mathcal{O} (z^2) ,
\end{equation}
so that, within this approximation, the expression \eqref{eq 0.25} can be rewritten in the form
\begin{align}
\mathcal{L}^{\left(1\right)} & =\frac{1}{\beta^{4}}\left[\frac{\pi^{2}}{45}+\frac{3}{16\pi^{2}} 
\left(2\beta m\right)^{2} \left( \frac{2}{\left(\beta m\right)^2} \underset{k=1}{\sum} \frac{1}{k^4}-\frac{1}{2}\underset{k=1}{\sum} \frac{1}{k^2} + \mathcal{O} (\left(k\beta m\right)^2) \right)
\right].\label{eq 0.25b}
\end{align}
 so the above sums can be written in terms of Riemann zeta function, so that
\begin{equation}
\mathcal{L}^{\left(1\right)}\simeq\frac{1}{\beta^{4}}\left[\frac{\pi^{2}}{45}+  \frac{ \pi^2}{60} - \frac{1}{16}    \left(\beta m\right)^{2}  \right].\label{eq 0.28}
\end{equation}
We then notice a correction due to GUP at the same order as in the (constant) coefficient Stefan-Boltzmann law. This will be further discussed later.
 

\section{Two-loop calculation}

\label{sec4}

\begin{figure*}[tbp]
\centering
\mbox{\subfigure []{ \epsfig{figure=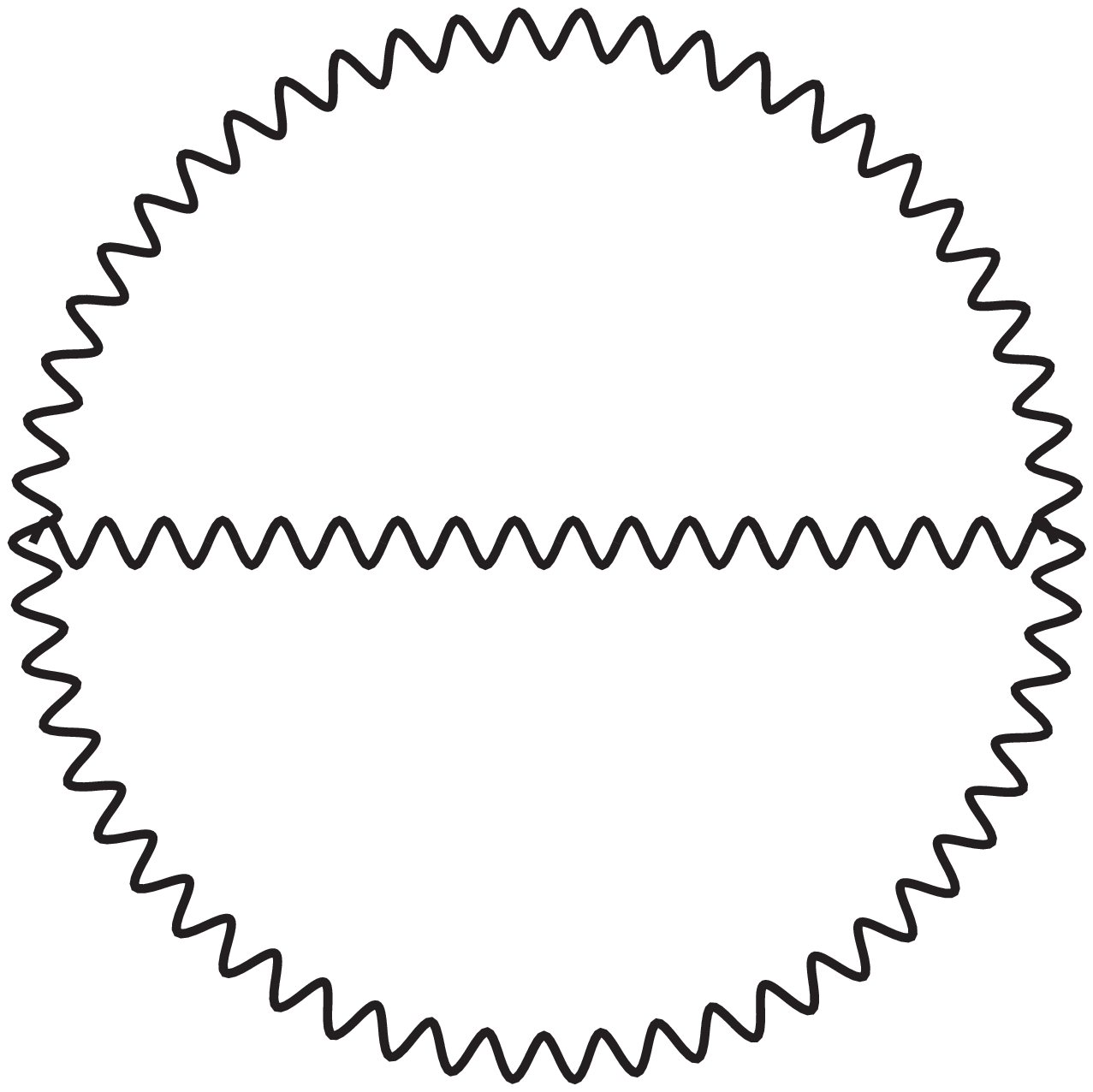,width=1.5in}}\quad
\subfigure []{\epsfig{figure=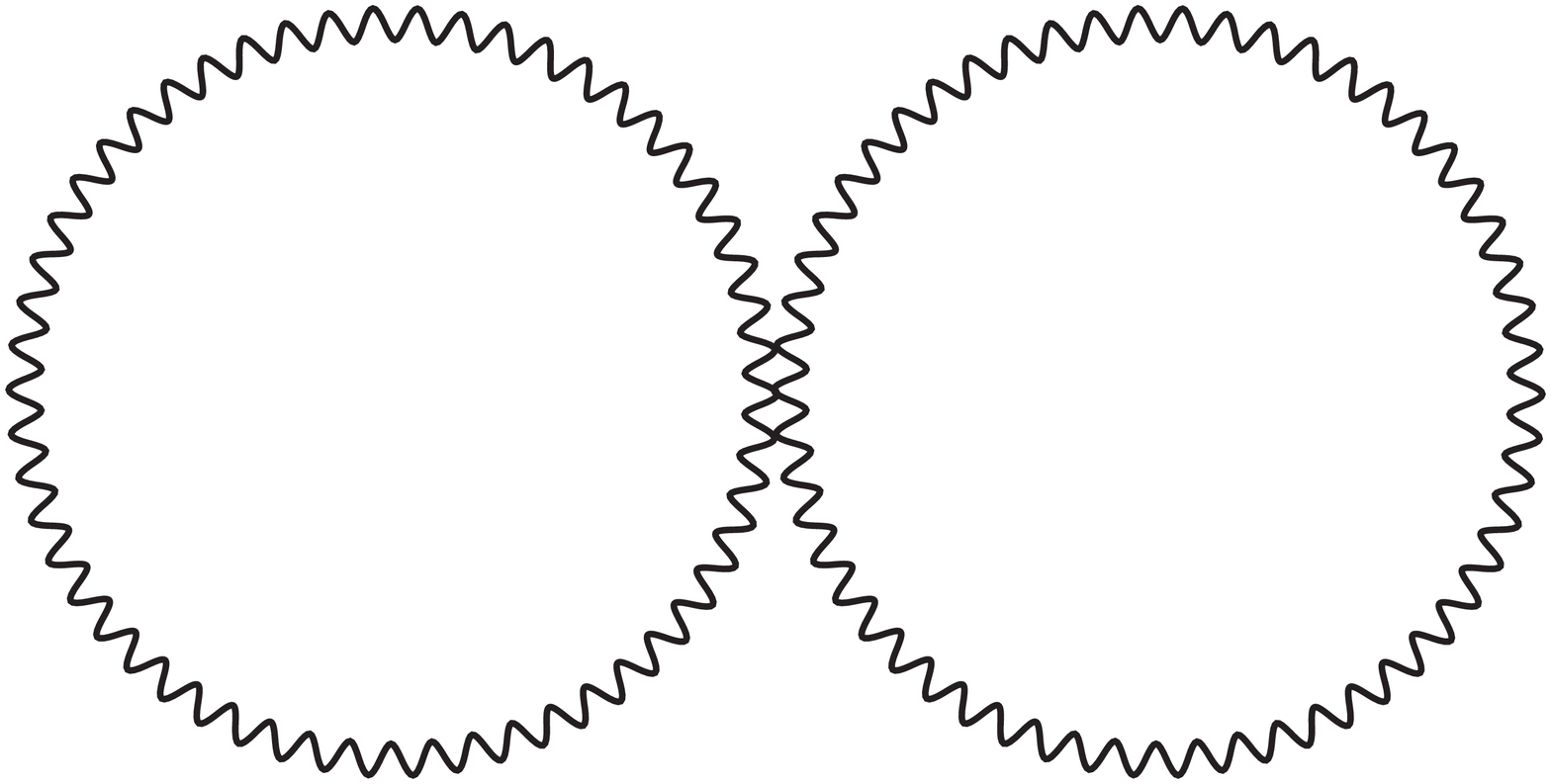,width=2.3in}}}
\caption{ The two-loop contributions coming from the three- and four-point functions.}
\label{graph2}
\end{figure*}

Since we have already shown the role played by the higher-derivative term by computing the one-loop contribution, we wish now to highlight the part played by the interactions induced by the generalized GUP. For this purpose we will now proceed and compute the two-loop order effective Lagrangian, the two contributing diagrams are shown at  Fig.~\ref{graph2}. The contribution (a) reads
\begin{align}
\mathcal{L}_{a}^{\left(2\right)}= & \frac{1}{2\beta^2} \underset{n_{B}}{\sumint  } \underset{m_{B}}{\sumint  }iD_{\mu\nu}\left(q\right)i\Gamma^{\mu\lambda\chi}\left(q,k,-p\right)iD_{\lambda\theta}\left(k\right)i\Gamma^{\phi\theta\nu}\left(p,-k,-q\right)iD_{\chi\phi}\left(p\right),\nonumber \\
= & \frac{1}{2\beta^{2}}\underset{n_{B}}{\sumint  } \underset{m_{B}}{\sumint  }\left[\frac{1}{k^{2}}-\frac{1}{k^{2}-m^{2}}\right]\left[\frac{1}{p^{2}}-\frac{1}{p^{2}-m^{2}}\right]\left[\frac{1}{q^{2}}-\frac{1}{q^{2}-m^{2}}\right]\nonumber \\
 & \times\left[\eta_{\mu\nu}\eta_{\lambda\theta}\eta_{\chi\phi}i\Gamma^{\mu\lambda\chi}\left(q,k,-p\right)i\Gamma^{\phi\theta\nu}\left(p,-k,-q\right)\right], \label{eq:4.1}
\end{align}
where $k=p-q$; while the contribution (b) is given as 
\begin{align}
\mathcal{L}_{b}^{\left(2\right)}= & \frac{1}{8\beta^2} \underset{n_{B}}{\sumint  } \underset{m_{B}}{\sumint  } iD_{\mu\nu}\left(p\right)i\Gamma^{\mu\nu\lambda\sigma}\left(p,-p,-q,q\right)iD_{\lambda\sigma}\left(q\right),\nonumber \\
= & \frac{1}{8\beta^{2}}\underset{n_{B}}{\sumint  } \underset{m_{B}}{\sumint  }  \left[\frac{1}{p^{2}}-\frac{1}{p^{2}-m^{2}}\right]\left[\frac{1}{q^{2}}-\frac{1}{\left(q-m^{2}\right)}\right]\left[\eta_{\mu\nu}\eta_{\lambda\sigma}i\Gamma^{\mu\nu\lambda\sigma}\left(p,-p,-q,q\right)\right]. \label{eq:4.2}
\end{align}

The contribution (a), Eq.~\eqref{eq:4.1}, is rather intricate, and after computing the tensor contraction and performing some simplifications 
we find
\begin{align}
\mathcal{L}_{a}^{\left(2\right)} & =-\frac{\alpha^{2}m^{6}}{8\beta^{2}} \underset{n_{B}}{\sumint  } \underset{m_{B}}{\sumint  }  \frac{1}{\left(k^{2}-m^{2}\right)}\frac{1}{\left(p^{2}-m^{2}\right)}\frac{1}{\left(q^{2}-m^{2}\right)}\nn
 & \times\biggl[\frac{2k^{2}}{q^{2}}+\frac{2p^{4}}{k^{2}q^{2}}-4-\frac{3}{2}\frac{k^{4}}{p^{2}q^{2}}-7\frac{p^{2}}{q^{2}}-\frac{2p^{2}}{k^{2}}\biggr]. \label{eq:4.3}
\end{align}
However, at finite temperature, the massive term is not easily handled, neither in order to get a closed expression for it, specially in the form present in Eq.~\eqref{eq:4.3}. Hence, as we have considered in the Sect.\ref{sec3}, we shall regard henceforth the (high-temperature) approximation $m^2/k^2 \approx \beta ^2 m^2 \gg 1$ in \eqref{eq:4.3}, which is consistent with the hard thermal loop approximation, and then consider the leading terms we are able to obtain
\begin{align}
\mathcal{L}_{a}^{\left(2\right)}  & \simeq-\frac{\alpha^{2}m^{4}}{8\beta^{2}} \left( \underset{n_{B}}{\sumint  } \biggl[4\frac{1}{ q^{2}-m^{2} }+\frac{1}{2}\frac{1}{q^{2}}\biggr]  \right)\left(\underset{m_{B}}{\sumint  }   \frac{1}{ p^{2}-m^{2} }\right). \label{eq:4.4}
\end{align}
We consider next the second contribution \eqref{eq:4.2}, the tensor contraction is readily computed and result into
\begin{align}
\mathcal{L}_{b}^{\left(2\right)} & =\frac{9\alpha m^{2}}{\beta^{2}} \left( \underset{n_{B}}{\sumint  } \biggl[\frac{1}{q^{2}}-\frac{1}{ q^{2}-m^{2} }\biggr]  \right)\left(\underset{m_{B}}{\sumint  }  \frac{1}{ p^{2}-m^{2} }  \right). \label{eq:4.5}
\end{align}
Hence, the sum of the two contributions Eqs.\eqref{eq:4.4} and \eqref{eq:4.5} gives the total two-loop contribution 
\begin{align}
\mathcal{L}^{\left(2\right)} & =\mathcal{L}_{a}^{\left(2\right)}+\mathcal{L}_{b}^{\left(2\right)}\\
   & =\frac{\alpha^2 m^4}{\beta^{2}} \left( \underset{n_{B}}{\sumint  }  \biggl[\frac{143}{16}\frac{1}{q^{2}}-\frac{19}{2}\frac{1}{ q^{2}-m^{2} }\biggr]\right)\left(\underset{m_{B}}{\sumint  }  \frac{1}{ p^{2}-m^{2} }\right),  \label{eq:4.11}
\end{align}
We can compute these sum/integration as follows: first the massless part that gives
contribution 
\begin{align}
\frac{1}{\beta}\underset{m_{B}}{\sumint  }\frac{1}{q^{2}} & =-\frac{1}{2\pi^{2}}\beta^{-2}\Gamma\left(2\right)\zeta\left(2\right)=-\frac{\beta^{-2}}{12}  \label{eq:4.10}
\end{align}
where we have made use of the integration  \eqref{eq 0.22} and the bosonic sum
\[
\frac{1}{\beta} \underset{m_{B}}{\sum}\frac{1}{\left(\frac{2\pi m_b}{\beta}\right)^2+\omega_q^2} = \frac{1}{\omega_q}\left[\frac{1}{2}+ \frac{1}{e^{\beta \omega_q}-1}\right]
\]
Now the massive integration requires further care in its evaluation
\begin{align}
\frac{1}{\beta}\underset{m_{B}}{\sumint  }\frac{1}{q^{2}-m^{2}} & =-\frac{1}{2\pi^{2}}\beta^{-2}\int dz\frac{z^{2}}{\sqrt{z^{2}+a^{2}}}\frac{1}{e^{\sqrt{z^{2}+a^{2}}}-1} \label{eq:4.6}
\end{align}
where we have defined $a^{2}=\beta^{2}m^{2}$. In order to gain insights
about the behavior of the above integral, we can consider the high-temperature
limit, so that we obtain approximately the leading value of the integral  \cite{Dolan:1973qd}.
In this case, we may consider the expansion \cite{Dolan:1973qd}
\begin{equation}
I\left(a^{2}\right) \equiv \int dz\frac{z^{2}}{\sqrt{z^{2}+a^{2}}}\frac{1}{e^{\sqrt{z^{2}+a^{2}}}-1}=\left.I\left(a^{2}\right)\right|_{a^{2}=0}+a^{2}\left.\frac{\partial I\left(a^{2}\right)}{\partial a^{2}}\right|_{a^{2}=0}+\frac{a^{4}}{2!}\left.\frac{\partial^{2}I\left(a^{2}\right)}{\partial a^{4}}\right|_{a^{2}=0}+...  \label{eq:4.7}
\end{equation}
the first term is well defined 
\begin{equation}
\left.I\left(a^{2}\right)\right|_{a^{2}=0}=\int dz\frac{z}{e^{z}-1}=\Gamma\left(2\right)\zeta\left(2\right)=\frac{\pi^{2}}{6}
\end{equation}
while the second term
\begin{align}
\frac{\partial I\left(a^{2}\right)}{\partial a^{2}} &  =-\int dz\frac{1}{\sqrt{z^{2}+a^{2}}}\frac{1}{e^{\sqrt{z^{2}+a^{2}}}-1}
\end{align}
however, demands further care, because the limit $a=0$
leads to a singular result. Hence, it is convenient to study the regulated
quantity 
\[
\frac{\partial I_{\epsilon}\left(a^{2}\right)}{\partial a^{2}}=-\int dz\frac{z^{-\epsilon}}{\sqrt{z^{2}+a^{2}}}\frac{1}{e^{\sqrt{z^{2}+a^{2}}}-1}
\]
with $0<\epsilon<1$. This regulated expression is known \cite{Dolan:1973qd} and can be straightforwardly computed yielding 
\begin{align}
\frac{\partial I_{\epsilon}\left(a^{2}\right)}{\partial a^{2}} &  =\frac{1}{2}\ln\frac{a}{4\pi}-\frac{\pi}{2a}-\frac{1}{2}\gamma+\mathcal{O}\left(a^{2}\right)+\mathcal{O}\left(\epsilon\right).
\end{align}
With these results we then obtain the following expression containing the leading terms of the expansion  \eqref{eq:4.7}
\begin{align}
\int dz\frac{z^{2}}{\sqrt{z^{2}+a^{2}}}\frac{1}{e^{\sqrt{z^{2}+a^{2}}}-1} & \simeq \frac{\pi^{2}}{6}+a^{2}\left(\frac{1}{2}\ln\frac{a}{4\pi}-\frac{\pi}{2a}-\frac{1}{2}\gamma\right)  \label{eq:4.8}
\end{align}
Hence, with the result \eqref{eq:4.8}, we find that the massive contribution \eqref{eq:4.6} reads
\begin{align}
\frac{1}{\beta}\underset{m_{B}}{\sumint  }\frac{1}{q^{2}-m^{2}}   & =-\frac{1}{12}\beta^{-2}-\frac{m^{2}}{8\pi^{2}}\ln\frac{\beta^{2}m^{2}}{4\pi}+\frac{1}{4\pi}\frac{m}{\beta}+\frac{m^{2}}{4\pi^{2}}\gamma  \label{eq:4.9}
\end{align}
Finally, replacing the obtained results \eqref{eq:4.9} and \eqref{eq:4.10} back into the expression \eqref{eq:4.11}, we obtain that the two-loop contribution to the effective action is
\begin{align}
\mathcal{L}^{\left(2\right)} & \simeq \frac{ \alpha^2 m^4 }{\beta^4} \biggl[-\frac{1}{256}+\frac{161}{768\pi}m\beta+\frac{1}{32\pi^{2}}\left(\frac{161\gamma}{24}-19\right)m^{2}\beta^{2}-\frac{19\gamma}{16\pi^{3}}m^{3}\beta^{3}-\frac{19\gamma^{2}}{32\pi^{4}}m^{4}\beta^{4}\nonumber \\
 & -\left( \frac{161}{1536\pi^{2}}m^{2}\beta^{2}-\frac{19}{32\pi^{3}}m^{3}\beta^{3}-\frac{19\gamma}{32\pi^{4}}m^{4}\beta^{4}\right)\ln\frac{\beta^{2}m^{2}}{4\pi} \biggr].  \label{eq:4.12}
\end{align}
Notice that the effective Lagrangian computed here is equal to $ \mathcal{L}_{\rm{eff}} =\ln Z$, thus we can determine any thermodynamical quantities.
Hence, to highlight the GUP effects from the ordinary behavior into obtained results it is useful to compute some of these quantities. In this way, we proceed in computing the internal energy density 
\begin{equation}
	u \left(T\right) = - \left(1+\beta\frac{d}{d\beta}\right) \mathcal{L}^{\rm{total}}
\end{equation}
where $\mathcal{L}^{\rm{total}}$ is the sum of the one- and two-loop contributions, Eqs. \eqref{eq 0.28} and \eqref{eq:4.12}, respectively.
Hence, performing the derivative of the above expression, we find
\begin{align}
u\left(T\right) & =\frac{\pi^{2}}{15}\frac{1}{\beta^{4}}+\left( \frac{\pi^2}{20}-\frac{1}{16}m^2 \beta^2 \right)\frac{1}{\beta^{4}} \nonumber \\
 & -\frac{\alpha^{2}m^{4}}{\beta^{4}}\biggl[\frac{3}{256}-\frac{161}{384\pi}m\beta-\frac{1}{768\pi^{2}}\left(161\gamma-295\right)m^{2}\beta^{2}+\frac{19}{16\pi^{3}}m^{3}\beta^{3}+\frac{19\gamma^{2}}{32\pi^{4}}m^{4}\beta^{4}\nonumber \\
 & +\left(\frac{161}{1536\pi^{2}}m^{2}\beta^{2}+\frac{19\gamma}{32\pi^{4}}m^{4}\beta^{4}\right)\ln\frac{\beta^{2}m^{2}}{4\pi}\biggr].  \label{eq:4.15}
\end{align}
It should be remarked that the first term in the right-hand side is the usual Stefan-Boltzmann law, $u = \sigma T^4$ where $\sigma = \pi^2/15$, and the remaining terms can be thought as corrections $\delta\sigma_{\rm{gup}}$ to the law due to GUP effects (even the constant terms is corrected in this case).
Moreover, equation  \eqref{eq:4.15} can be used in the description of new phenomena that involve both massless and massive propagating modes for the gauge field.  
For instance, it has recently been proposed that a nonvanishing photon mass  ($m_\gamma \leq 10^{-27}eV$)
can be used rather than a cosmological constant ($\Lambda \sim m_\gamma^2$) to explain dark energy consistent with the current observations \cite{Kouwn:2015cdw}.
In this cosmological scenario thermal effects of massive photons as described here could have prominent role.


\section{Concluding remarks}
\label{sec5}

In this paper we have considered the thermodynamics of a photon gas subject to a deformed Heisenberg algebra, or more precisely with the presence of a minimal measurable length.
The analysis followed a field theoretical point-of-view in order to compute the effective Lagrangian (partition function).
In particular, we have made use of a proposed covariant extension of the original generalized uncertainty principle.
After a brief review of this extension, we proceed in order to determine a dynamics for the photon field. For that matter,
we first defined a fermionic matter action and then by resorting to local gauge invariance, we introduced a GUP covariant derivative $\mathcal{D}_{\mu}$ that transforms correctly under the given local transformation, i.e. $\mathcal{D}_{\mu}\rightarrow U\left(x\right)\mathcal{D}_{\mu}U^{\dagger}\left(x\right)$.
With this new GUP covariant derivative is straightforward to compute the field strength tensor by the usual identity $i\mathcal{F}_{\mu\nu}\Phi=\left[\mathcal{D}_{\mu},\mathcal{D}_{\nu}\right]\Phi$.
Finally, with this quantity, we can compute generalized invariants such as $\mathcal{F}_{\mu\nu} \mathcal{F}^{\mu\nu}$
and $\mathcal{F}_{\mu\nu} \mathcal{G}^{\mu\nu}$.
It is important to remark that we have taken an expansion in the minimal length $\alpha$ and considered $\mathcal{O}(\alpha^2)$ terms in our analysis.

After establishing a GUP modified Maxwell action, in which three- and four-point couplings are now present, we have computed the propagators and the respective vertex functions. 
Notice that the coefficient of these vertex functions are corrected by nonlocal contributions, i.e. higher-order contributions of the expansion in $\alpha$, so that the quantities computed here are basically the first-order approximation.
In order to highlight the GUP effects we wish to compute thermodynamical quantities. In this way, we considered the one- and two-loop order contribution to the effective Lagrangian (partition function) at the high-temperature limit.
Thus, the obtained additional terms can be seen as corrections $\delta\sigma_{\rm{GUP}}$ to the Stefan-Boltzmann law due to GUP effects.

Since we have found the presence of a higher-derivative term in the deformed action, we might also wish to circumvent this illness by considering a different covariant algebra, in which the temporal coordinates are as the usual, and the higher-derivatives are only present at the spatial coordinates, i.e. $\hat{x}^{\mu}=x^{\mu},\quad\hat{p}_{\nu}=\left(p_{0},\left(1+\alpha\vec{p}^{2}\right)p_{i}\right)$. This can be regarded as a Horava-Lifshitz-like theory, since there are no ghosts (negative energy modes) present. Moreover, we can proceed as before and define a covariant derivative such as $\mathcal{D}_{\mu}=\left(\nabla_{0},\left(1-\alpha\nabla_{k}\nabla_{k}\right)\nabla_{i}\right)$ in order to perform an analysis of this Horava-Lifshitz-like field theory. The subject is under consideration and will be reported elsewhere.
 
 \subsection*{Acknowledgements}
 
 The author would like to thanks the anonymous referee
for his/her comments and suggestions to improve this
paper.



\end{document}